\documentstyle[11pt]{article} 
\newcommand{\tmiv}{10^{-4}}
\newcommand{\tmv}{10^{-5}}

\newcommand{\zb}{Z}
\newcommand{\wb}{W}            
\newcommand{\mz}{M_{_Z}}
\newcommand{\mw}{M_{_W}}
\newcommand{\mh}{M_{_H}}
\newcommand{\mzs}{M^2_{_Z}}
\newcommand{\seffsf}[1]{\sin^2\theta^{#1}_{\rm{eff}}}
\newcommand{\afba}[1]{A^{#1}_{_{\rm FB}}}
\newcommand{\alra}[1]{A^{#1}_{_{\rm LR}}}
\newcommand{\baru}{\overline u}
\newcommand{\ord}[1]{{\cal O}\lpar#1\rpar}
\newcommand{\mt}{m_t}
\newcommand{\mts}{m^2_t}
\def\alr{A_{_{\rm LR}}}
\newcommand{\sigh}{\sigma_{\rm had}}
\newcommand{\gn}{\Gamma_{\nu}}
\newcommand{\gel}{\Gamma_{e}}
\newcommand{\gmu}{\Gamma_{\mu}}

\newcommand{\gt}{\Gamma_{\tau}}

\newcommand{\gu}{\Gamma_{u}}
\newcommand{\gd}{\Gamma_{d}}
\newcommand{\gc}{\Gamma_{c}}

\newcommand{\gbq}{\Gamma_{b}}
\newcommand{\gz}{\Gamma_{_Z}}
\newcommand{\gh}{\Gamma_{h}}
\newcommand{\gi}{\Gamma_{\rm{inv}}}
\newcommand{\bq}{\begin{equation}}                    
\newcommand{\eq}{\end{equation}}
\newcommand{\tbn}[1]{Table \ref{#1}}
\newcommand{\tbns}[2]{Tables \ref{#1}--\ref{#2}}

\newcommand{\lpar}{\left(}                            
\newcommand{\rpar}{\right)}

\newcommand{\sla}[1]{/\!\!\!#1}
\newcommand{\gfd}{\gamma_5}                    
\newcommand{\siws}{s^2_{_W}}
\newcommand{\GeV}{\;\mathrm{GeV}}
\newcommand{\MeV}{\;\mathrm{MeV}}

\def\gappeq{\mathrel{\rlap {\raise.5ex\hbox{$>$}}
{\lower.5ex\hbox{$\sim$}}}}

\def\lappeq{\mathrel{\rlap{\raise.5ex\hbox{$<$}}
{\lower.5ex\hbox{$\sim$}}}}

\begin{document}

\pagestyle{empty}
\begin{flushright}
{CERN-TH/98-92}\\
{\tt hep-ph/9803425}
\end{flushright}
\vspace*{10mm}
\begin{center}
{\bf UPGRADING OF PRECISION CALCULATIONS\\
 FOR ELECTROWEAK OBSERVABLES}
 \\
\vspace*{15mm} 
{\bf Dmitri Bardin$^{1}$\vspace{0.5cm}\\
and\vspace{0.5cm}\\
Giampiero Passarino$^{2}$}\\
\vspace{15mm}
$^1$ Laboratory of Nuclear Problems, JINR, Dubna, Russia$^{*}$  
\\[5mm]
$^2$ Dipartimento di Fisica Teorica, Universit\`a di Torino and \\
     INFN, Sezione di Torino, Turin, Italy$^{\dagger}$
\vspace*{2cm}  

{\bf Abstract} 
\\ 
\end{center}
\normalsize 
\noindent A critical assessment is given of the 
comparison between the new versions of the programs TOPAZ0 40i and 
ZFITTER 510. 
The relevance for precision calculations around the $\zb$ resonance is
briefly discussed.
\vspace*{1.5cm}

\noindent 
\rule[.1in]{16.5cm}{.002in}

\noindent
$^{*)}$ Present address: TH Division, CERN, 1211 Geneva 23, Switzerland.\\
e-mail address: Dmitri Bardine@cern.ch. \\
$^{\dagger)}$ e-mail address: giampiero@to.infn.it.
\vspace*{1.0cm}

\begin{flushleft} CERN-TH/98-92 \\
March 1998 
\end{flushleft}
\vfill\eject

\setcounter{page}{1}
\pagestyle{plain}

\vspace{2cm}

\def\thefootnote{\arabic{footnote}} \setcounter{footnote}{3}

\setcounter{page}{1}

\section{Introduction}

In 1995 the $\zb$ phase of LEP came to an end and at present the ultimate 
analysis of the data is imminent. This involves in particular the completion 
of the line-shape analysis, including the final LEP energy calibration. 
Consequently, the safest possible estimate of the theoretical accuracy is of 
the utmost importance. It should be noted that the LEP\,1 data (1990--1995) 
were taken in the energy ($\sqrt{s}\,$) range 
$| \sqrt{s} - \mz | < 3\;$GeV and consist of the hadronic 
and leptonic cross sections, the leptonic forward--backward asymmetries, the 
various polarization asymmetries, the partial widths, and the quark 
forward--backward asymmetries. All this makes it mandatory to assess the 
theoretical precision of the available programs for different channels and for 
different pseudo-observables.

In this note we focus on the calculation of the pseudo-observables. 
Independently of the renormalization procedure that is used, the matrix
element for $\zb \to f{\bar f}$ will be written as
\bq
{\cal M}^{\zb}_{f{\bar f}} = \baru_f \sla{e_{_Z}}
\lpar {\cal G}^f_{_V} + {\cal G}^f_{_A}\,\gfd\rpar v_f.
\eq
With the above results we can now define the {\em pseudo-observable}
quantities that are relevant for the phenomenology of LEP~1/SLC.
The {\em pseudo-variables} are related to measured cross sections and
asymmetries by some deconvolution or {\it unfolding} procedure.
The concept of {\it pseudo-observability} is introduced by saying that the 
experiments {\em measure} some primordial (basically cross sections and 
thereby asymmetries also) quantities, which are then reduced to secondary 
quantities under some set of specific assumptions. Within these assumptions 
the secondary quantities, the {\em pseudo-observables}, also deserve the label 
of {\em observability}.

In 1995 the CERN Report~\cite{yr95} on `Precision Calculations for the $\zb$ 
resonance' provided as basic documentation the theoretical basis for upgrading 
the '89 Report on $\zb$ Physics at LEP~1'~\cite{yr89}.

Although the previous analyses remain quite comprehensive, an update of the 
discussion of radiative corrections has become necessary for one very good
reason: a sizeable amount of theoretical work has appeared following the
CERN report of 1995. In particular, a crucial amount of work has been
performed in providing higher-order QCD corrections, mixed electroweak and QCD
corrections~\cite{mix}, and sub-leading two-loop corrections of 
$\ord{\alpha^2\mts}$~\cite{dfs}.

In ref.~\cite{dfs} the two-loop $\ord{\alpha^2\mts}$ corrections are 
incorporated in the
theoretical calculation of $\mw$ and $\seffsf{l}$. More recently
the complete calculation of the decay rate of the $\zb$ has been made 
available to us~\cite{prel}. 
The only case that is not covered is the one of final 
$b$-quarks, because it involves specific $\ord{\alpha^2\mts}$ vertex 
corrections.

Another recent development in the computation of radiative corrections to the 
hadronic decay of the $\zb$ is contained in two papers, which, together, 
provide
complete corrections of $\ord{\alpha\alpha_s}$ to $\Gamma(\zb\to q\bar{q})$
with $q=u,d,s,c$ and $b$.
In the first reference of~\cite{mix} the decay into light quarks is treated. 
In the second one the remaining diagrams contributing to the decay into 
bottom quarks are considered and thus the mixed two-loop corrections 
are complete.

\section{Numerical results and comparison}
Two of the programs described in the '95 CERN Report have been constantly
updated~\cite{updated},
and we focus, in this note, on a comparison between TOPAZ0~\cite{topaz0} and
ZFITTER~\cite{zfitter}, with an update of the predictions of 
$\zb$-resonance observables 
within the minimal standard model.

\begin{table}[t]
\begin{center}
\begin{tabular}{|c||c|c|c|c|}
  \hline
  & \multicolumn{4}{c|}{NEW versions} \\
  \cline{2-5}
  & TOPAZ0 & ZFITTER & Rel. dev. (per mille) & Abs. dev.  \\
  \hline \hline
  \multicolumn{5}{|c|}{$\mh = 100\,$GeV} \\
  \hline \hline
$\mw\,$[GeV]  & 80.3864   & 80.3860   &  0.005  &  0.4   [MeV] \\
$\gn\,$[MeV]  & 167.235   & 167.262   & --0.16   & --0.027 [MeV] \\
$\gel\,$[MeV] & 84.0028   & 84.0140   & --0.13   & --0.011 [MeV] \\
$\gmu\,$[MeV] & 84.0021   & 84.0133   & --0.13   & --0.011 [MeV] \\
$\gt\,$[MeV]  & 83.8110   & 83.8237   & --0.15   & --0.013 [MeV] \\ 
$\gu\,$[MeV]  & 300.372   & 300.387   & --0.050  & --0.015 [MeV] \\
$\gd\,$[MeV]  & 383.161   & 383.187   & --0.068  & --0.026 [MeV] \\
$\gc\,$[MeV]  & 300.315   & 300.329   & --0.047  & --0.014 [MeV] \\
$\gbq\,$[MeV] & 376.100   & 376.082   &  0.048  &  0.018 [MeV] \\
$\gz\,$[MeV]  & 2496.62   & 2496.81   & --0.076  & --0.19  [MeV] \\
$\gh\,$[MeV]  & 1743.10   & 1743.17   & --0.040  & --0.07  [MeV] \\
$\gi\,$[MeV]  & 501.706   &  501.787  & --0.16   & --0.081 [MeV] \\
$\seffsf{l}$  & 0.231489  & 0.231516  & --0.12   & --0.000027    \\
$\seffsf{b}$  & 0.232788  & 0.232902  & --0.49   & --0.000114    \\
$\afba{l}$    & 0.0162774 & 0.0162225 &   -     &  0.0000549   \\
$\afba{b}$    & 0.103232  & 0.103094  &   -     &  0.000138    \\
$\afba{c}$    & 0.0738276 & 0.0736708 &   -     &  0.0001568   \\
$\alr$        & 0.147320  & 0.147071  &   -     &  0.000249    \\
$\sigh\,$[nb] & 41.4717   & 41.4734   & --0.041  & --0.0017 [nb] \\
$R_l$         & 20.7505   & 20.7486   &  0.092  &  0.0019      \\
$R_b$         & 0.215765  & 0.215746  &  0.088  &  0.000019    \\
$R_c$         & 0.172288  & 0.172289  & --0.006  & --0.000001    \\
$\alra{b}$    & 0.934703  & 0.934638  &  0.070  &  0.000065    \\
$\alra{c}$    & 0.667961  & 0.667892  &  0.103  &  0.000069    \\
$\siws$       & 0.222855  & 0.222862  &  0.031  & --0.000007    \\
  \hline \hline
  \hline 
\end{tabular}
\end{center}
\caption[]{Comparison of TOPAZ0 40i and ZFITTER 510 for $\mh = 100\,$GeV. 
Here $\mz = 91.1867\,$GeV, $\mt = 175.6\,$GeV, $\alpha_s(\mzs) = 0.120$,  
$1/\alpha_{\rm{em}}(\mzs)=128.896$.} 
\label{comparison100}
\end{table}

In \tbn{comparison100} we compare the prediction of TOPAZ0 and ZFITTER for
$\mz = 91.1867\,$GeV, $\mt = 175.6\,$GeV, $\alpha_s(\mzs) = 0.120$ and
$\mh = 100\,$GeV. The results are from the new versions of the programs,
and we have also shown absolute and relative (in per mille) deviations 
for a set
of $25$ pseudo-observables. The relative deviation is defined as
\bq
\delta = 2\,{{\rm TOPAZ0 - ZFITTER}\over {\rm TOPAZ0 + ZFITTER}}.
\eq
For quantities such as the asymmetries, we report the absolute deviation,
which is the only relevant one.

\begin{table}[t]
\begin{center}
\begin{tabular}{|c||c|c|c|c|}
  \hline
  & \multicolumn{4}{c|}{NEW versions} \\
  \cline{2-5}
  & TOPAZ0 & ZFITTER & Rel. dev. (per mille) & Abs. dev.  \\
  \hline \hline
  \multicolumn{5}{|c|}{$\mh = 200\,$GeV} \\
  \hline \hline
$\mw\,$[GeV]  &  80.3417  &  80.3416  &  0.001  &  0.1\,[MeV]  \\
$\gn\,$[MeV]  &  167.181  & 167.198   & --0.10   & --0.017 [MeV] \\
$\gel\,$[MeV] & 83.9576   & 83.9635   & --0.070  & --0.006 [MeV] \\
$\gmu\,$[MeV] & 83.9569   & 83.9629   & --0.071  & --0.006 [MeV] \\
$\gt\,$[MeV]  & 83.7658   & 83.7733   & --0.090  & --0.008 [MeV] \\
$\gu\,$[MeV]  & 300.083   & 300.080   &  0.010  &  0.003 [MeV] \\
$\gd\,$[MeV]  & 382.863   & 382.866   & --0.008  & --0.003 [MeV] \\
$\gc\,$[MeV]  & 300.026   & 300.022   &  0.013  &  0.004 [MeV] \\
$\gbq\,$[MeV] & 375.785   & 375.785   &   0     &   0    [MeV] \\
$\gz\,$[MeV]  & 2494.83   & 2494.91   & --0.032  & --0.08  [MeV] \\
$\gh\,$[MeV]  & 1741.61   & 1741.62   & --0.006  & --0.01 [MeV]  \\
$\gi\,$[MeV]  & 501.542   & 501.593   & --0.10   & --0.051 [MeV] \\
$\seffsf{l}$  & 0.231849  & 0.231872  & --0.099  & --0.000023    \\
$\seffsf{b}$  & 0.233150  & 0.233247  & --0.42   &  0.000097    \\
$\afba{l}$    & 0.0156576 & 0.0156088 &   -     &  0.000488    \\     
$\afba{b}$    & 0.101185  & 0.101101  &   -     &  0.000084    \\
$\afba{c}$    & 0.0722734 & 0.0721303 &   -     &  0.0001431   \\
$\alr$        & 0.144488  & 0.144263  &   -     &  0.000225    \\     
$\sigh\,$[nb] & 41.4734   & 41.4746   & --0.029  & --0.0012 [nb] \\
$R_l$         & 20.7439   & 20.7426   &  0.063  &  0.0013      \\
$R_b$         & 0.215769  & 0.215768  &  0.005  &  0.000001    \\
$R_c$         & 0.172269  & 0.172266  &  0.017  &  0.000003    \\
$\alra{b}$    & 0.934471  & 0.934416  &  0.059  &  0.000055    \\
$\alra{c}$    & 0.666715  & 0.666657  &  0.087  &  0.000058    \\
$\siws$       & 0.223719  & 0.223721  &  0.009  & --0.000002    \\
  \hline \hline
  \hline 
\end{tabular}
\end{center}
\caption[]{Comparison of TOPAZ0 40i and ZFITTER 510 for $\mh = 200\,$GeV. 
Here $\mz = 91.1867\,$GeV, $\mt = 175.6\,$GeV, $\alpha_s(\mzs) = 0.120$, 
$1/\alpha_{\rm{em}}(\mzs)=128.896$.} 
\label{comparison200}
\end{table}

A similar comparison for $\mh = 200\,$GeV is shown in \tbn{comparison200}.
We observe a deviation in $\seffsf{l}$ of $2.7\times 10^{-5}\;(2.3\times
10^{-5})$ for $\mh = 100\,\GeV\;(200\,\GeV)$. For the total $\zb$ width, the
difference between the two programs is of $0.19\,\MeV\;(0.08\,\MeV)$ for
$\mh = 100\,\GeV\;(200\,\GeV)$. 

In the hadronic $\zb$ width the difference
is $0.07\,$MeV at $\mh = 100\,$GeV, which roughly corresponds to a variation
of $\Delta\alpha_s(\mzs) = 0.00013$ in the predictions for $\alpha_s(\mzs)$
from the two programs. Variations for the $\wb$ mass are everywhere 
below $0.5\,$MeV.

The level of agreement that is reached is highly satisfactory, especially 
if we take into account the fact  
that the implementation of the new correction factors has
been performed in a completely independent way, different renormalization 
schemes and, more important, absolutely different strategies.

\begin{table}[t]
\begin{center}
\begin{tabular}{|c||c|c|c|c|}
\hline
& \multicolumn{4}{c|}{NEW versus OLD} \\
  \cline{2-5}
& TOPAZ0 & ZFITTER & Rel. dev. (per mille) & Abs. dev.  \\
  \hline \hline
  \multicolumn{5}{|c|}{$\mw\,$[GeV]} \\
\hline \hline
Old &  80.310  &  80.317  &   --0.09   &  --7 [MeV]     \\
New &  80.308  &  80.308  &   --0.01   &  --0.1 [MeV]   \\
\hline
Rel. shift(per mille) &  --0.03          &  --0.11       & &  \\
Abs. shift          &  --2.1 [MeV]     &  --9.0 [MeV]  & &  \\
  \hline \hline
  \multicolumn{5}{|c|}{$\gel\,$[MeV]} \\
  \hline \hline
Old &  83.931  &  83.941  &   --0.12   &  --0.01 [MeV]  \\
New &  83.915  &  83.920  &   --0.07   &  --0.006 [MeV]  \\
\hline
Rel. shift(per mille) &  --0.19          & --0.25        & &  \\
Abs. shift          & --0.016 [MeV]    & --0.021 [MeV] & &  \\
  \hline \hline
  \multicolumn{5}{|c|}{$\seffsf{l}$} \\
  \hline \hline
Old &  0.23200 &  0.23205 &  --0.22    & --5.0$\times\tmv$  \\
New &  0.23209 &  0.23211 &  --0.09    & --2.0$\times\tmv$  \\
\hline
Rel. shift(per mille) &  0.39           &  0.26        & &  \\
Abs. shift          & 8.9$\times\tmv$ & 6.1$\times \tmv$  & &\\
  \hline \hline
  \multicolumn{5}{|c|}{$\afba{l}$} \\
  \hline \hline
Old & 0.015360 & 0.015310 &     -     &  5.0$\times \tmv$  \\
New & 0.015249 & 0.015204 &     -     &  4.5$\times \tmv$  \\
\hline
Abs. shift     &  --1.1$\times \tmiv$  & --1.1$\times\tmiv$ & & \\
  \hline \hline
  \multicolumn{5}{|c|}{$\alr$} \\
  \hline \hline
Old &  0.14327 &  0.14289 &     --     &  3.8$\times \tmiv$ \\
New &  0.14259 &  0.14238 &     --     &  2.1$\times \tmiv$ \\
\hline
Abs. shift     & --6.8$\times \tmiv$   & --5.1$\times\tmiv$ & & \\
  \hline \hline
  \multicolumn{5}{|c|}{$\gz\,$[MeV]} \\
  \hline \hline
Old &  2497.4  &  2497.5  &  --0.04    &   --0.1 [MeV]     \\
New &  2496.1  &  2496.2  &  --0.04    &   --0.1 [MeV]     \\
\hline
Rel. shift(per mille) &  --0.52          &   --0.52           & & \\
Abs. shift          &  --1.29 [MeV]    &   --1.30 [MeV]     & & \\
  \hline \hline
  \hline 
\end{tabular}
\end{center}
\caption[]{Comparison of TOPAZ0 and ZFITTER. 
Here $\mz = 91.1888\,$GeV, $\mt = 175\,$GeV,
$\mh = 300\,$GeV, $\alpha_s(\mzs) = 0.125$ and
$1/\alpha_{\rm{em}}(\mzs)=128.896$.} 
\label{oldnew1}
\end{table}
\begin{table}[t]
\begin{center}
\begin{tabular}{|c||c|c|c|c|}
  \hline
  & \multicolumn{4}{c|}{NEW versus OLD} \\
  \cline{2-5}
  & TOPAZ0 & ZFITTER & Rel. dev. (per mille) & Abs. dev.  \\
  \hline \hline
  \multicolumn{5}{|c|}{$R_l$} \\
  \hline \hline
Old &  20.782  &  20.781  &    0.05   &    0.001  \\
New &  20.773  &  20.773  &    0      &    0      \\
\hline
Rel. shift(per mille) &    --0.41        &  --0.39          & &  \\
Abs. shift          &    --0.0085      &  --0.008         & &  \\
  \hline \hline
  \multicolumn{5}{|c|}{$R_b$} \\
  \hline \hline
Old & 0.21567  &  0.21571 &   --0.19   &   --4.0$\times \tmv$  \\
New & 0.21579  &  0.21580 &   --0.05   &   --1.0$\times \tmv$  \\
\hline
Rel. shift(per mille) &    0.57         &   0.42          & &  \\
Abs shift           & 1.2$\times\tmiv$& 9.0$\times\tmv$ & &  \\
  \hline \hline
  \multicolumn{5}{|c|}{$R_c$} \\
  \hline \hline
Old &  0.17237 &  0.17236 &   0.06    &   1.0$\times \tmv$   \\
New &  0.17235 &  0.17235 &    0      &         0            \\
\hline
Rel. shift(per mille) &  --0.09          &  --0.06          & &  \\
Abs. shift          &--1.6$\times\tmv$ & --1.0$\times\tmv$& & \\
  \hline \hline
  \multicolumn{5}{|c|}{$\afba{b}$} \\
  \hline \hline
Old & 0.10033  & 0.10013  &     -     &   2.0$\times \tmiv$  \\
New & 0.099815 & 0.099767 &     -     &   4.8$\times \tmv$   \\
\hline
Abs. shift          &--5.1$\times\tmiv$&--3.6$\times\tmiv$& &  \\
  \hline \hline
  \multicolumn{5}{|c|}{$\afba{c}$} \\
  \hline \hline
Old & 0.071590 & 0.071380 &     -     &  2.1$\times \tmiv$   \\
New & 0.071235 & 0.071100 &     -     &  1.3$\times \tmiv$   \\
\hline
Abs. shift &   --3.6$\times \tmiv$ &   --2.8$\times \tmiv$ & & \\
  \hline \hline
  \hline 
\end{tabular}
\end{center}
\caption[]{Comparison of TOPAZ0 and ZFITTER. 
Here $\mz = 91.1888\,$GeV, $\mt = 175\,$GeV, 
$\mh = 300\,$GeV, $\alpha_s(\mzs) = 0.125$ and
$1/\alpha_{\rm{em}}(\mzs)=128.896$.} 
\label{oldnew2}
\end{table}
It is also interesting to compare the present situation with the differences
registered between the two codes at the time of the '95 CERN Report.
For this reason we have taken again $\mz = 91.1888\,$GeV, $\mt = 175\,$GeV,
$\alpha_s(\mzs) = 0.125$ and $\mh = 300\,$GeV (the '95 input parameter
set) and compared some of the predictions. In \tbns{oldnew1}{oldnew2} we give 
the comparison showing, at the same time, the {\em old--old} and
{\em new--new} deviations and the shifts {\em old--new}.

It is worth noticing that {\em new--new} deviations are always less (or much 
less) than the corresponding {\em old--old} ones. This fact induces a sizeable
reduction of the theoretical uncertainty, achieved after the implementation of 
the new correction factors. We have also compared our results without the 
sub-leading terms $\ord{\alpha^2\mts}$ and found again the same level of
agreement as reached in '95.

In conclusion, we have achieved an important goal: after a substantial
upgrading, TOPAZ0 and ZFITTER continue to agree with each other extremely well,
in most cases better than they ever have. 

An important consequence of
this fact is that the central value of the Higgs boson mass in the famous 
$\chi^2(\mh)$ curve moves down from 115 GeV
obtained with old versions to approximately $87\,$GeV.
The
difference in predictions between TOPAZ0 and ZFITTER is less than
$5\,$GeV in any of the fits performed so far~\cite{LEPEWWG}.
Morever, a substantial reduction is expected of the entire {\em blue band} 
that is giving the theoretical uncertainty in the same curve.
\section{Aknowledgements}

We would like to express special thanks to the TOPAZ0 and ZFITTER 
teams~\cite{teams}. 
Without their contributions the two programs would not be what they are.
We both would like to express special thanks to Martin Gr\"unewald, 
Hans K\"uhn, Christoph Pauss, and G\"unter Quast.   
Finally we acknowledge the important role played by Giuseppe Degrassi and by
Paolo Gambino in helping us with the implementation of the two-loop
sub-leadings corrections and for sharing with us the result of their work
prior to publication.


\end{document}